\documentclass[12pt,preprint]{aastex}
\shorttitle{Near-IR spectra of young stars in Aquila}
\shortauthors{Rice et al.}
\usepackage{epsfig}
\begin{document}

\title{An Association in the Aquila Star-Forming Region: \\ High
  Resolution Infrared Spectroscopy of T~Tauri Stars}

\author{E. L. RICE\altaffilmark{1}, L. PRATO\altaffilmark{2}, AND
  I. S. MCLEAN\altaffilmark{1} }

\altaffiltext{1}{Department of Physics and Astronomy, UCLA,
Los Angeles, CA 90095; erice@astro.ucla.edu}

\altaffiltext{2}{Lowell Observatory, 1400 West Mars Hill
  Rd. Flagstaff, AZ 86001}

\begin{abstract}
We present the properties of a group of young stars associated
with the well-studied T~Tauri star system AS~353, located in the
Aquila star-forming region.  The association is identified using radial velocity measurements of
sample objects selected from the Herbig and Bell Catalog based on
their spatial proximity to AS~353.  Radial velocities of nine objects were measured from multi-epoch
high-resolution (R$\sim$30,000) $H$-band spectra obtained with NIRSPEC
on Keck~II.~ 
High-resolution $K$-band spectra were also obtained for most of the sample objects.
Spectral types and rotational velocities are determined for all objects in the sample.
The multi-epoch $H$-band spectra were examined for radial velocity
variations in order to detect possible spectroscopic binaries.
Eight of the nine objects have radial velocities that are consistent
within the 1-$\sigma$ scatter of the sample.  
From their mean of $-$8~km~s$^{-1}$ these eight objects have a
standard deviation of 2~km~s$^{-1}$, which suggests that the sample stars are related.
The ninth object shows significant radial velocity variations between
epochs, characteristic of a spectroscopic binary.  The overall multiplicity of the sample
is high; we observed 13 stars in seven systems, identifying three new
candidate binary components in this project.
Many of the spectra reveal hydrogen emission lines typical of strong accretion processes,
indicating that most of these objects harbor circumstellar disks and are less than a few
million years old.  Based on previous estimates, we adopt a distance
of 200$\pm$30~pc to the young stars in Aquila in order to calculate
luminosities and place the stars on an H-R diagram.  We discuss possible
interpretations of the enigmatic pure emission line spectrum of
HBC~684.  This work represents the highest spectral
resolution infrared observations to date of these intriguing, nearby young stars.

\end{abstract}


\section{\sc Introduction}

Despite much study of the formation and evolution of young T~Tauri
stars, many questions remain concerning some of their most fundamental characteristics.  
Understanding the few million years of evolution between the
protostellar and post T~Tauri stages, including circumstellar disk
accretion, planet formation, and disk dissipation, is crucial to
understanding the formation of the Solar System and extra-solar planetary systems. 
One key approach to resolving these issues is striking a balance
between in-depth studies of individual star-forming regions and
observations of a variety of star-forming regions that differ in age,
distance, stellar density, binary fraction, etc.    
The presence of T~Tauri stars in a region indicates an age of 1$-$10
million years, the timescale for disk dissipation and the formation of
gas giant planets around low mass stars \citep[e.g.,][]{Zuc01}.
T~Tauri stars are primarily identified by H$\alpha$ emission, IR and
UV continuum excess, variability, and surrounding nebulosity \citep[e.g.,][]{Whi01,Zuc04}. 

After 50 years of study, the young, low mass multiple star system AS~353,
in the constellation Aquila, remains enigmatic.  A $\sim$5\arcsec~pair, the system
appears to be a mixed T~Tauri binary, with the more massive component, AS~353A,
showing signatures of a classical T~Tauri star (cTTs), which is defined
as having EW(H$\alpha$)~$>$~10~\AA~and excess infrared emission
above the continuum level \citep{Str89}. The secondary, AS~353B, was classified
as a weak-lined T~Tauri star (wTTs, EW(H$\alpha$)~$<$~10~\AA), but has recently
been resolved into a subarcsecond binary \citep{Tok04}.  AS~353Ba was
reclassified by \citet{Tok04} as a cTTs (but see
\S4.5.1 for further discussion).  AS~353 is the
exciting source of Herbig-Haro object HH~32,  a highly collimated bipolar
outflow \citep[see][and references therein]{Cur97} and one of the only Herbig-Haro
objects detected in 3.6 and 6~cm radio continuum \citep{Ang92,Ang98}.  
Radio continuum emission has also been detected from the stars
themselves, most recently attributed to
the secondary rather than the primary \citep{Avi01}.

The age of the AS~353 system was most recently determined by
\citet{Pra03} based on the location of AS~353B relative to the
pre$-$main-sequence evolutionary tracks of \citet{Pal99}.  
This analysis resulted in an age of 5$\times$10$^{5}$ years\footnote{An error appears
in Figure 7 of \citet{Pra03} because of a missing minus sign in the Log (L$_*$/L$_{\odot}$)
for AS~353B.  Data in the tables and text are accurate.}, or 1$\times$10$^{6}$
years if the binary nature of AS~353B is taken into account (see
Appendix B of \citealt{Pra03}).  In this current paper we
provide an updated discussion of the age of the AS~353 system as well
as best estimates for the other objects in our sample (\S 4.2).

In addition to AS~353B, there are five T~Tauri stars listed in the
Herbig and Bell Catalog (1988, hereafter HBC) that are within
$\sim$15\degr~of AS~353A, all between galactic latitudes of $b$~=~$-$1\degr~to~$-$6\degr.  
Their projected location in the crowded galactic plane complicates the
study of these stars, as well as the search for additional young stars in the region.
Few of the HBC Aquila objects have radial velocities or $v{\sin}i$
listed in the catalog, but from spectra with
resolution lower than the observations presented here and/or using
emission lines instead of absorption lines, with results that have
uncertainties on the order of tens of km~s$^{-1}$ or more \citep{Her77,Eis90,Fer95}.

The original motivation for this project was to search for radial
velocity variations in a sample of cTTs and wTTs in the vicinity of
AS~353A as a means of identifying spectroscopic binaries.
For double-lined spectroscopic binary systems, the relative velocity
of each component, the mass ratio, and the center of velocity of the
system can be determined.  Obtaining dynamical masses of young stars is critical to the study of
star formation because mass is the most fundamental of all stellar
parameters, and evolutionary models that relate young star masses and
ages with the observables L and T$_{eff}$ are not yet calibrated.
Furthermore, absolute radial velocities are important for determining
if the young stars in the sample are actually related, i.e. if they
were formed in the same molecular cloud.
If they were, then it is expected that they have a low velocity
dispersion relative to one another as well as with respect to the
associated cloud \citep[e.g.,][]{Her77}.

High resolution infrared spectra, even a single echelle order, can reveal much
more than radial velocities.
In this study, what were assumed to be typical classical or weak-lined T~Tauri stars
were found to be surprisingly active systems with high rotational
velocities, strong veiling, and unusual emission lines.
Thus the goals of this project broadened to include the
measurement of $v{\sin}i$, determination of spectral type, and
characterization of the circumstellar environments of the targets, in
addition to the radial velocity measurements.
The sample is introduced in \S2.1, and the NIRSPEC observations and
their reduction are described in \S\S2.2 and 2.3, respectively.
The data analysis, including line identification,
determination of spectral type and $v{\sin}i$, and measurements of
radial velocity variability and absolute radial velocity, are
described in\S3.  The results are discussed in \S4,
including details of circumstellar environment, age, membership, and multiplicity.
A summary appears in \S5 along with suggestions for further studies of
the Aquila star-forming region.

\section{\sc Observations \& Data Reduction}

\subsection{Sample}

The targets were selected from the Third Catalog of Emission-Line
Stars of the Orion Population (HBC) because of their spatial proximity
to the well-studied T~Tauri star AS~353A (HBC~292) and its companion, AS~353B (HBC~685). 
The names, coordinates, $H$-band magnitudes, and equivalent width of H$\alpha$ emission for
each object are listed in Table~\ref{table:coords}. 
Coordinates are from SIMBAD\footnote{SIMBAD is provided by the Centre
  de Donn\'ees astronomiques de Strasbourg, France,
  http://simbad.u-strasbg.fr/Simbad.} unless otherwise noted.
The objects all lie within a region from 19h~00m to 19h~40m right
ascension and $-$5\degr~to~$+$11\degr~declination (Equinox 2000.0),
placing them in the constellation Aquila.
We adopt a distance of 200~pc~$\pm$~30~pc (\S4.2), which is consistent with both
the most recent distance estimate to AS~353 (150$\pm$50~pc,
\citealt{Pra03}) and to the dark cloud complex associated with HBC~294
(230$\pm$30, \citealt{Kaw01}).  The physical size of this region on the plane of the
sky is estimated to be 32$\pm$5~pc by 58$\pm$9~pc at this distance.
Targets with previously determined spectral types range from early K to early M
\citep{Coh79}; several have been revised as a result of the work described in this paper.

As originally selected the sample consisted of six objects: HBC 292, 294, 681, 682, 683, and 684.  
During the course of the observations and subsequent analysis, the sample has grown to nine objects. 
HBC~682 was discovered to be a 1\arcsec~visual binary upon the first
observation with NIRSPEC.  AS~353B was added to the sample in order to strengthen its
association, using radial velocity measurements, with AS~353A.
Because of its close proximity to HBC~681 and 682, a ninth star,
FG~Aql/G3 \citep{Coh79}, was also observed.

\subsection{Observations}

Data were obtained with the cryogenic, cross-dispersed echelle
spectrometer NIRSPEC installed on the Nasmyth platform of the Keck~II
telescope at the W.M.~Keck Observatory on Mauna Kea, Hawaii \citep{McL98,McL00}.   
NIRSPEC employs a 1024 $\times$ 1024 ALADDIN InSb array detector
operating at 30~K and sensitive to the range 0.95 to 5.4~$\micron$.
All observations were taken in high-resolution mode with a slit size
of 0.288\arcsec$\times$24\arcsec~for non-AO observations and
0.027\arcsec$\times$2.26\arcsec~for AO observations.
The echelle mode and two pixel slit width provided a resolving power
(R=$\lambda$/$\Delta\lambda$) of R$\sim$30,000 for non-AO
observations and R$\sim$35,000 for observations behind AO,
corresponding to a velocity resolution of $\sim$10~km~s$^{-1}$ per
two-pixel resolution element in non-AO mode and $\sim$8.6~km~s$^{-1}$ in AO mode.
Integration times ranged from two to five minutes per exposure for the Aquila objects.
For each observation, a series of four spectra was taken in an ABBA
pattern, nodding $\sim$10\arcsec~along the slit between the A and B positions.
Typical seeing was 0\farcs6.  Multiple ABBA sets were taken for
fainter objects.  Integration times ranged from 120~s to 300~s (Table~\ref{table:obslog}).

$H$-band observations using the N5 filter (1.413~-~1.808~$\micron$)
were taken at several epochs between 2002 July
and 2004 November to search for radial velocity variations.
The intention was to observe each object a minimum of three times,
however, it was possible to observe HBC~683 only twice and
AS~353B and FG~Aql/G3 only once.  
On 2004 May 27 and 28 (UT) the observations were made using NIRSPEC
behind the Keck Adaptive Optics system \citep{Wiz00}.

A $K$-band spectrum was obtained for each object on either 2003
September 07 (HBC numbers 294, 681, 682A\&B, 683, and 684) or 2004
November 22 (AS~353A).
AS~353B and FG~Aql/G3 were not observed in $K$-band. 
Slightly different echelle angles resulted in different spectral coverage for
the two sets of $K$-band observations.  $K$-band spectra obtained on
2003 September 07 (not shown) suffered from poor subtraction of telluric
absorption lines.
Table~\ref{table:obslog} provides an observing log; instrument
settings for all observations are summarized in Table~\ref{table:inst}. 

\subsection{Data Reduction}

For data reduction we used the REDSPEC
code,\footnote{http://www2.keck.hawaii.edu/inst/nirspec/redspec/} a
package of IDL procedures created specifically for the reduction
of NIRSPEC spectra by S. Kim, L. Prato, and I. McLean.
Median-filtered dark and flat fields were created from multiple
flat lamp and dark exposures for non-AO $H$-band observations.
For AO observations and $K$-band observations, only a single
flat and dark exposure were taken and differenced
to create a final flat field.
We determined the dispersion solution with a target star frame using
OH night sky emission lines \citep{Rou00} or an arc lamp frame.
The latter approach was necessary for the AO data because the night
sky OH emission lines are too faint in the target frames.
Based on the error analysis reported in \citet{Pra02}, we assign
an uncertainty of 1~km~s$^{-1}$ to the dispersion of each target.

We used $H$-band order 49 (1.545~-~1.567~$\micron$) for the radial
velocity measurements because of the presence of many narrow atomic and molecular
absorption lines and the lack of telluric absorption lines.
Therefore, no division by a telluric A0 star was needed for order 49 spectra.
In our $K$-band analysis we focused on the Br$\gamma$ line in order 35
(2.159~-~2.190~$\micron$); orders 33, 36 and 37 were also reduced.

Reduction of order 35 was complicated by the presence of the broad
Br$\gamma$ absorption in the telluric A0 star spectrum.
To remove this intrinsic spectral feature while retaining the telluric
features, we reduced the order 35 A0 star spectrum leaving the broad
Br$\gamma$ and narrow terrestrial features intact.
In a second reduction, we interpolated over the narrow telluric absorption
lines only, leaving the Br$\gamma$ feature intact.
The A0 star spectra resulting from these two reductions were then subtracted
and the difference spectrum was renormalized to a continuum of 1.0,
creating a spectrum of telluric features only.
Target spectra were then divided by this corrected A0 star spectrum to
remove the telluric lines.

The spectra were transformed onto a heliocentric reference frame and
the continua were flattened and normalized to a value of 1.0. 
The signal to noise ratio (SNR) of the spectra was estimated by
determining the maximum peak-to-peak variation in an order and dividing
this by a factor of 4 to estimate the noise.  The
SNR appears in the last column of Table~\ref{table:obslog}.

\section{\sc Results and Analysis}

\subsection{High-Resolution Spectra}

\begin{figure*}
\includegraphics[angle=0,width=5.0in]{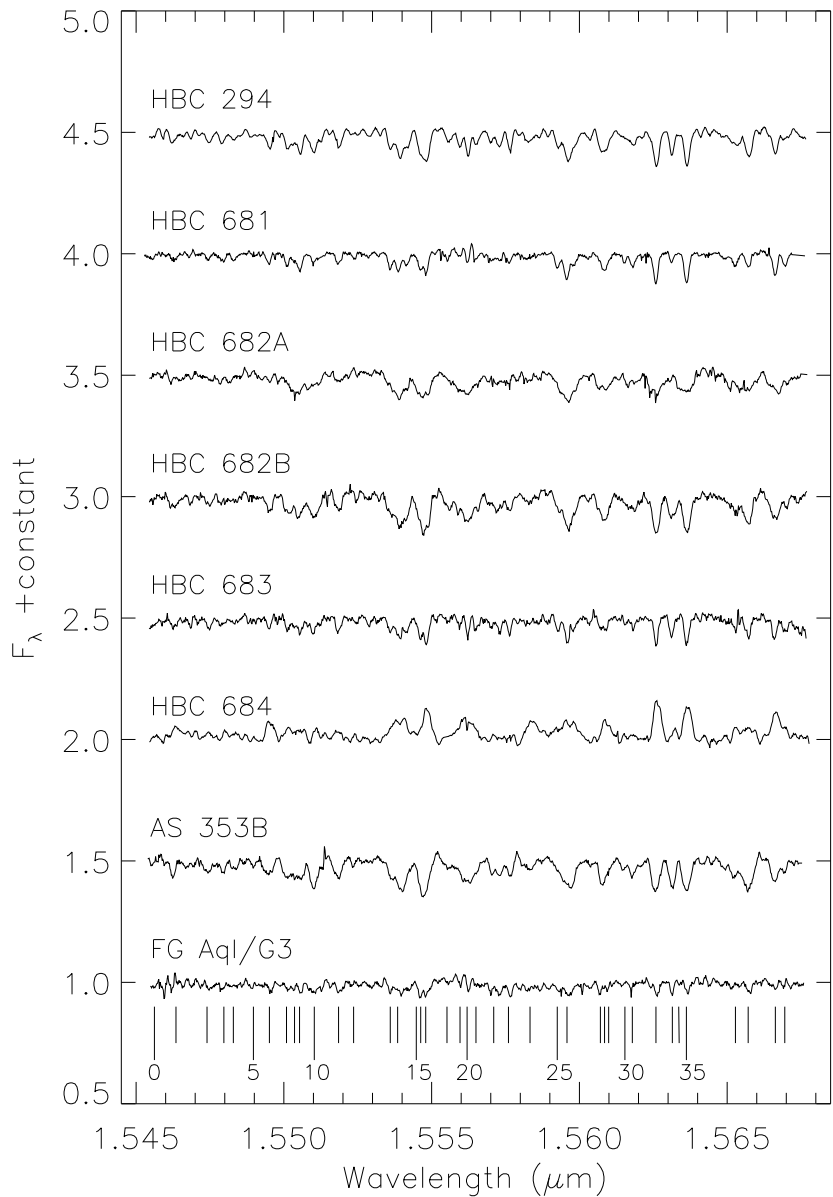}
\caption{NIRSPEC order 49 spectra of eight sample objects. All spectra
  are from observations made on 2003 September 08 (UT) except for AS~353B
  (2002 July 18) and FG~Aql/G3 (2003 August 10).  Wavelength and
  species of identified lines are given in Table~\ref{table:lines}.  All spectra
  have been wavelength-corrected for heliocentric and radial velocity,
  normalized to a continuum level of unity, and offset by 0.5 in flux. }
\label{fig:08sep03}
\end{figure*}

Figures~\ref{fig:08sep03}~through~\ref{fig:hbc684} present the
high-resolution spectra that will be discussed further in \S4.
Absorption and emission lines were identified using
atlases of the solar spectrum and sunspot umbral spectrum in the
near-infrared \citep{Liv91,Wal01}.
$K$-band line identifications are from Table~4 of \citet{Pra03}.
Individual objects are discussed in detail in \S4.5.

Almost 40 absorption lines (some blended) were identified
in the order 49 spectra of the object with the
lowest rotational velocity, HBC~683, and therefore the sharpest lines.
Fewer individual lines are identified in spectra with higher
rotational velocity because nearby lines become increasingly blended as $v{\sin}i$ increases.
The vast majority of the lines are atomic: mainly neutral Fe, some
neutral Ti, Ni, and Si, and possibly S and C are identified.
Intrinsic OH absorption appears in most spectra and a possible
detection of CN emission in HBC~684.
Significant $H$-band veiling in all of the absorption line object
spectra is evident when these are compared to template spectra at similar $v{\sin}i$.
Strong emission lines of very different nature were detected in order
49 for two objects: Br16 in AS~353A and multiple atomic, and
possible molecular, lines in HBC~684.

In the $K$-band spectra for the sample, neutral atomic lines, such as
Na, He, Al, and Mg, are observed in orders 33, 34, 36, and 37 (e.g.,
Figure~\ref{fig:as353a_k}).  Br $\gamma$ appears in order 35.  Molecular lines of
CO (orders 32 and 33) and H$_2$ (order 36) are also present.

The equivalent widths (EW) of the Br$\gamma$ line (n=7$-$4,
$\lambda$=2.1661~$\micron$)  were calculated for all objects observed
in the $K$-band.  For AS~353A, the Br$\gamma$~EW, $-22\pm$1~\AA, is
consistent with the Br$\gamma$ EW of $-21\pm1$~\AA~from \citet{Pra03},
derived from low spectral resolution (R$\sim$800) observations made
in 1996.  EW was also calculated for the Br16 (n=16$-$4,
$\lambda$=1.5661~$\micron$) line observed in $H$-band order~49 of AS~353A.
Line fluxes were calculated using the equivalent widths and $K$-band
magnitude from \emph{2MASS}.  
The uncertainty of the EW calculations is approximately $\pm$1~\AA,
corresponding to an uncertainty of
$\lesssim$~0.5~W$-$m$^{-2}\times10^{-16}$ for the Br$\gamma$ line
fluxes and $\sim$0.3~W$-$m$^{-2}\times10^{-16}$ for the Br16 line fluxes.
Brackett line EWs and fluxes appear in Table~\ref{table:EWs}.

\subsection{Spectral Types and $v{\sin}i$}

The lines that were primarily used to determine spectral type
were two Fe lines at 1.56259~$\micron$ and 1.56362~$\micron$ and a
stellar OH doublet between them at 1.56314~$\micron$ in order 49.
For a K1 star, the Fe lines are prominent and the OH line is nearly absent.
For later K spectral types, the Fe lines monotonically weaken while the
OH line strengthens; all three lines are of approximately equal strength for spectral type M1.

\citet{Ben05} convolved rotation profiles from \citet{Gra92} with the
spectra of observed spectral type standard stars, resulting in
template spectra, for a range of spectral types,
with $v{\sin}i$=0, 2, 4, 6, 8, 10, 12, 15, 20, 25, 30, 40, and 50~km~s$^{-1}$.
Examples of the template spectra with $v{\sin}i$=10~km~s$^{-1}$
are presented in Figure~\ref{fig:templates}; these objects
are listed in Table~\ref{table:templates}.  Uncertainties in the
radial velocities of the template spectra are on the order of
1~km~s$^{-1}$ \citep{Pra02}.
The best fit spectral types were estimated by comparison
with these standard star template spectra from \citet{Ben05}
and then confirmed with cross-correlation.
The spectral types are accurate to at least two
subclasses; there are templates for about every second
subclass, but the difference between the template spectra is generally
well-defined, especially in the Fe and OH lines mentioned above.

Because of veiling of the object spectra by continuum emission
from reprocessing in circumstellar disks and accretion-induced excesses,
cross-correlation yields an artificial best fit for high
values of $v{\sin}i$.  Therefore, the $v{\sin}i$ of each object was 
determined by visual inspection and comparison to rotated
spectral templates alone.  The spectral type and $v{\sin}i$
for each object in the sample appear in Table~\ref{table:results}.

\subsection{Radial Velocities}

Radial velocities were measured by
cross-correlating the target spectra with the best fit
template spectrum from \citet{Ben05}.  Spectra of HBC~684 were
inverted because the typical absorption lines appear in emission
for this object (see \S4.5.5).  Before cross-correlation, spectra were
converted to the heliocentric reference frame,
interpolated onto the same wavelength basis as the matching
template spectrum, and rebinned by a factor of ten.

Because of the strong Br16 emission line in order 49, a different approach
was used to measure the radial velocity of AS~353A.
Absorption lines are more reliable for the measurement of radial
velocity of an object because emission lines are produced by
accretion flows or stellar winds, which have their own characteristic
radial velocities.  Therefore, we avoided the strong Br16 emission line in
the spectra of AS~353A and used only the absorption lines shortward and
longward of Br16 in the cross-correlation.

To estimate the uncertainties in the radial velocity measurements, the
target spectra were cross-correlated with five template spectra of
varying rotational velocity and spectral type, including the best-fit
template.  The standard deviation of the resulting radial velocity
measurements was less than 2~km~s$^{-1}$ for all targets other than
HBC~682A and HBC~684.  For several targets the standard deviation of
measurements was less than 1~km~s$^{-1}$.  We therefore adopt an
uncertainty in the measurement of 2~km~s$^{-1}$ for all targets.  This is consistent with the
more formal uncertainties for the same method derived by \citet{Pra02}
and \citet{Maz02}.

We measured each object's radial velocity at every epoch observed.
Table~\ref{table:radvels} lists the objects in column (1), the UT dates
of observation in column (2), and the radial velocities measured for
each object at each epoch in column (3).  For each object, we
averaged the radial velocities over all epochs.  These average values
are reported in column (4) of Table~\ref{table:results}.
The standard deviation and the number of epochs for each target 
appear in columns (5) and (6) of Table~\ref{table:results}.
No average radial velocity is reported for HBC~682A because of the
large variations between epochs. 
Standard deviations are not given for FG~Aql/G3 and HBC~685 because
only one observation of each object was made. 

Targets were initially examined for radial velocity variability by
cross-correlating spectra taken at different epochs.
For all objects except HBC~682A, radial velocity variations were less
than $\sim$4~km~s$^{-1}$, with typical values around 1$-$2~km~s$^{-1}$.
This is within the estimated uncertainty of the radial velocity measurements.
Radial velocity variations for HBC~682A were as large as 12~km~s$^{-1}$.
Velocity shifts between spectra from AO and non-AO observations were
systematically higher than the typical velocity shifts between non-AO
spectra from different epochs.  This may be the result of the different
approach to the dispersion solution necessary with AO observations.
The OH night sky lines used for the non-AO dispersion are distributed
regularly throughout order 49, allowing for an excellent solution.
The arc lamp lines required for the AO observations' dispersion solution
are only four and their distribution across order 49 is sparse.

\section{\sc Discussion}

\subsection{Circumstellar Properties}

To determine the general circumstellar properties of the sample we plot
$J-H$ versus $H-K$ colors from the \emph{2MASS} photometry (Figure~\ref{fig:ccd}).
Five of the objects have infrared excess according to this analysis,
consistent with the same five objects that have EW(H$\alpha$)~$>$~10~\AA~(Table~\ref{table:coords}).
Visual extinctions were estimated by dereddening the objects to the
cTTs locus \citep{Mey97} using

\begin{equation}
A_{v} = 13.83(J - H)_{obs} - 8.29(H - K)_{obs} - 7.43
\end{equation}

\noindent
from \citet{Pra03}.
The \emph{2MASS} colors and the derived extinctions appear in
Table~\ref{table:colors}.
The average visual extinction of the sample is 1.6 magnitudes, not
including AS~353A, which falls below the cTTs locus, possibly
as the result of scattered light or a blue component to the 
color from continuum excess emission produced in accretion flows.

\subsection{Age}

To estimate the ages of the Aquila stars, we place all of the
sample objects on an H-R diagram with the pre$-$main-sequence evolutionary
tracks of \citet{Pal99} (Figure~\ref{fig:hrd}).
Luminosities were calculated using the method described in \S3.3 of
\citet{Pra03}.  \emph{2MASS} photometry was used for all objects
  except HBC~294, for which the angularly resolved photometry of
  \citet{Age94} was used.  $J$-band bolometric corrections were from
  \citet[Table~4 in][]{Har94}.  Temperatures were taken from
  \citet{Luh04} based on the spectral types listed in Table~\ref{table:results}.  
Our adopted distance, 200~pc, was selected based on previous distance
estimates to stars in this region.  \citet{Pra03} used 150$\pm$50~pc
as the distance to AS~353B, citing distance estimates to the Aquila
Rift and Gould's Belt.  \citet{Kaw01} estimated the distance to the
dark cloud L694 to be 230$\pm$30~pc based on star counts.  We adopt a
distance that is consistent with both of these estimates, 200$\pm$30~pc.
The bolometric luminosities for the components of HBC~682 have been
corrected for binarity based on their near-infrared flux ratio of 2.25
and 2.15, calculated from $H$- and $K$-band unpublished NIRC2 images,
respectively.  For AS~353B the flux ratio of the components is $\sim$1, and for HBC~682
the flux ratio was calculated in $H$- and $K$-band from unpublished
NIRC2 images.

Based on a distance of 150~pc, the age of AS~353B was most
recently determined to be 5$\times$10$^{5}$ years \citep{Pra03}.
However, this calculation assumed that AS~353B was a single star;
\citet{Tok04} has since identified the system as a close binary
comprised of two M1.5 stars with a flux ratio of $\sim$1.  In placing
AS~353B on the HR diagram, we adopt our spectral type of M0
(Table~\ref{table:results} and \S4.5.1), a distance of 200~pc, and one-half the luminosity
determined from the unresolved system.  Our resulting improved age for
each component of AS~353B is $\sim$1~Myr.

Although the vertical error bars are large because of the uncertain
distances to the objects, the location of the ensemble on the H-R
diagram suggests that most of the objects are approximately 1~Myr
old.  The large scatter in age could be attributed to the uncertainty
in extinction down to the photospheres as well as uncertainty in the
amount of infrared excess, both of which could lead to over- or
under-estimation of bolometric luminosity for a particular object.
Disk signatures in the young set of objects are consistent with an age
of less than a few Myr. HBC 682A and B and especially FG~Aql/G3 appear
older, $\sim$6$-$12~Myr, consistent with their lack of emission lines
and IR excesses, although the spectrum of FG~Aql/G3 appears heavily
veiled (Figure~\ref{fig:08sep03}).

\subsection{Membership of the Aquila Association}

The standard deviation of the radial velocities of all the stars in
the sample ($\sim$2~km~s$^{-1}$), combined with many of the objects'
signatures of youth and spatial proximity, are consistent with most if
not all of the targets being members of a young stellar association.
The average radial velocity for each object ranges
from $-$6.8 to $-$11.4~km~s$^{-1}$; all but two of the objects have
radial velocities within one standard deviation of the mean of $-$8.6~km~s$^{-1}$.  

The only object for which a relatively constant radial velocity was not
measured is HBC~682A, indicating that it may be a spectroscopic binary.
Higher sampling frequency observations are needed to determine its mass
ratio and and center of mass velocity.  However, we speculate that 
HBC~682A is a probable member of the Aquila
association because of its proximity to HBC~682B ($\sim$1\arcsec).
Thus, it is likely that all nine stars in the sample belong to a
single association that is part of the Aquila star-forming region.

\subsection{Multiplicity}

The multiplicity among the Aquila association members is comparable to
that of Taurus, the well-studied star-forming region with the highest binary
fraction \citep[e.g.][]{Sim95,Sim97}.  Of seven systems that we observed, as many as thirteen
components were found.  Systems HBC~681 and HBC~682 were identified as
multiple candidates for the first time.  Two systems were known
binaries (HBC~294 and HBC~682) and at least one is triple (AS~353).
Additionally, HBC~682A may be a spectroscopic binary (\S4.5.3) and HBC~681 may
have a low-mass companion (\S4.5.2).

\subsection{Notes on Individual Objects}

\subsubsection{AS~353A \& B (HBC~292 \& HBC~685)}

\begin{figure*}
\includegraphics[angle=90,width=6.0in]{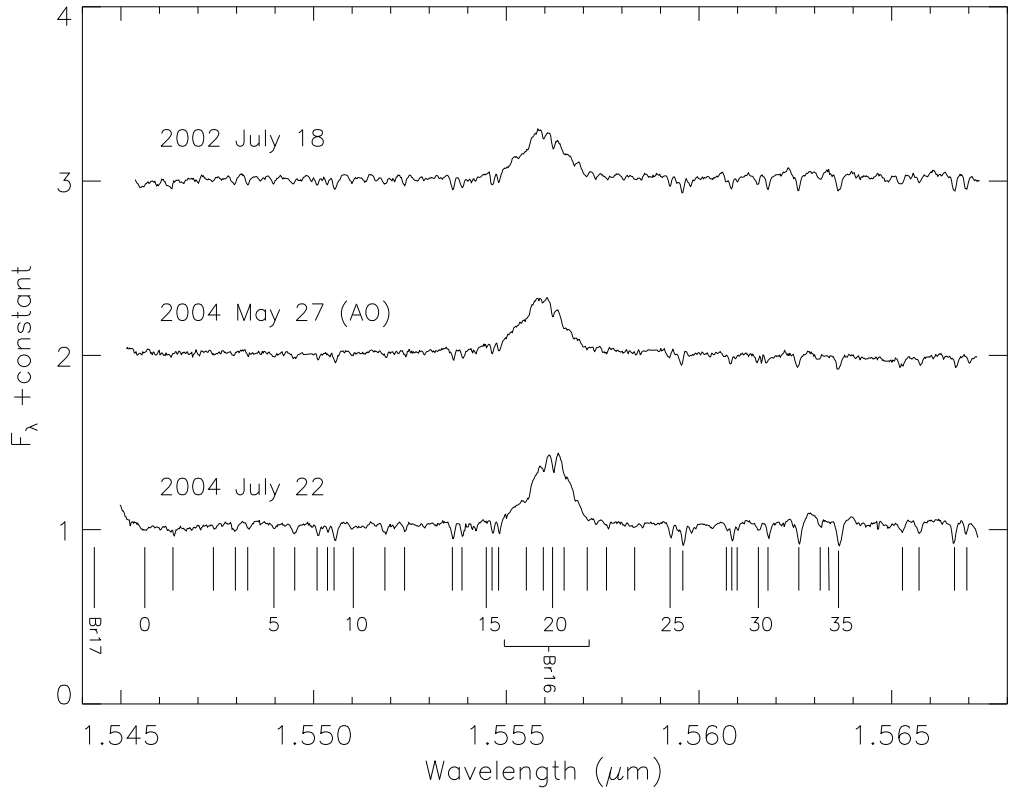}
\caption{Multi-epoch NIRSPEC order 49 spectra of the cTTs AS~353A labeled with UT observation dates.
  The prominent emission feature is Br16 (1.5561~$\micron$).  Wavelength and
  species of identified lines are given in Table~\ref{table:lines}.
  All spectra have been wavelength-corrected for heliocentric and
  radial velocity and normalized to a continuum level of unity.  }
\label{fig:as353a}
\end{figure*}

\begin{figure*}
\includegraphics[angle=0,width=5.0in]{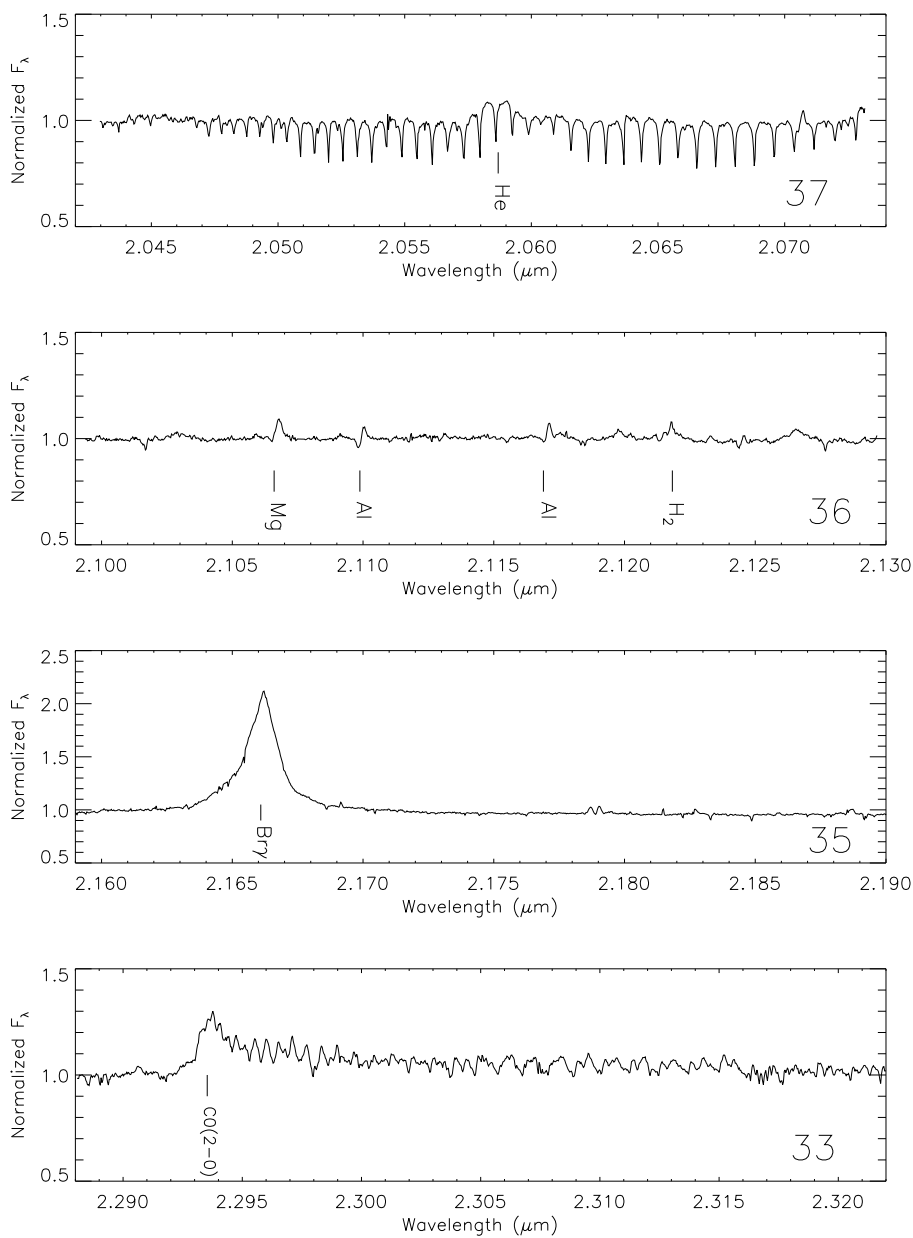}
\caption{Multi-order NIRSPEC $K$-band spectra of the cTTs AS~353A.
  All spectra have been normalized to a continuum level of unity within the order
  and wavelength-corrected for heliocentric and radial velocity.  Note
  that the F$_{\lambda}$ range is larger for order 35 in order to show
  the entire Br$\gamma$ emission line.  The series of evenly spaced
  absorption lines in order~37 is the residue of saturated telluric
  absorption lines.}
\label{fig:as353a_k}
\end{figure*}

The order 49 spectra from the three observations of AS~353A are
presented in Figure~\ref{fig:as353a}.  AS~353A is best fit
by a K5 template spectrum with $v{\sin}i$=10~km~s$^{-1}$, somewhat
later than the K2 spectral type previously reported in the literature \citep{Tok04}. 
The order 49 spectrum from the observation of AS~353B presented in
Figure~\ref{fig:08sep03} is best fit by an M0 template
spectrum with $v{\sin}i$=20~km~s$^{-1}$, which is similar to the M1.5
spectral types found for both components of
this subarcsecond binary by \citet{Tok04} using low resolution $K$-band spectra.

Strong Br16 (n=16$-$4) emission dominates the spectrum
of AS~353A in order 49.  A rise at the short wavelength end of
the 2004 July 22 (UT) spectrum also suggests the
presence of Br17, although the majority of the
line falls outside the spectral coverage of the order.
These strong Brackett lines are consistent with
the strength of the H$\alpha$ detected in the optical \citep{Her88}
and confirmed by us in a visible light echelle spectrum (not shown).
However, it is unusual to see what are presumably photospheric
absorption lines superimposed on the emission line.  The emission
line appears to shift relative to the absorption lines between epochs.
The veiling also appears variable; for example, the
absorption lines are markedly deeper in the UT 2004 July 22
spectrum than in earlier spectra (Figure~\ref{fig:as353a}).

Selected orders of the $K$-band spectra of AS~353A along with probable
feature identifications are presented in Figure~\ref{fig:as353a_k}.
Order 35 shows strong Br$\gamma$ emission, which
is again consistent with the strong H$\alpha$ emission.
Also apparent in the $K$-band spectra are atomic Mg, Al, and He, the
CO(2$-$0) bandhead, and possibly H$_2$, all in emission.  This rich
emission spectrum is also apparent at low resolution (R$\sim$760)
\citep{Pra03} and in the optical \citep{Eis90}.

Both AS~353A and B occupy unusual locations on the color-color diagram.
The position of AS~353A indicates significant IR excess,
consistent with the origin of the strong hydrogen emission in a
circumstellar disk. The detection of AS~353 by the \emph{IRAS}
satellite (\emph{IRAS}~19181$+$1056; the
components are unresolved in the \emph{IRAS} beam) also points to the
presence of warm dust in the system \citep{Wea92}.  However,
AS~353A lies {\it below} the cTTs locus, which may be the result
of abundant scattered light in the circumstellar environment or a
blue component of excess continuum emission from accretion flows.
AS~353B does not show IR excess on the color-color diagram, which is
consistent with its classification as a wTTs based on the strength of its H$\alpha$ emission.  
However, the extinction is larger than that of other cluster members
if it is dereddened to the cTTs locus.

The prominent neutral hydrogen emission from AS~353A
is likely a combination of accretion and stellar winds.
H$\alpha$ emission is the primary characteristic used to identify circumstellar accretion \citep{Whi01}.
However, a strong stellar wind is indicated by other characteristics
of this system, namely, that it is the exciting source of HH~32, a
system of bright, compact emission line regions that are produced by a
highly collimated bi-polar outflow \citep{Cur97}.
The H$\alpha$ emission from AS~353A also displays a
P~Cygni profile with absorption in the blue wing of the emission line
\citep{Fer95}, which indicates outflow from the star.

The similar radial velocity measurements of AS~353A and B ($-$11.4 and
$-$10.7~km~s$^{-1}$, respectively) and similar proper motions measured
by \citet{Her83} make it likely that they are indeed a bound pair.
Therefore, AS~353 is a mixed T~Tauri system in which the more
massive component is a cTTs while the secondary is a wTTs, which is
thought to be more evolved than a cTTs.  The slight infrared excess
implied by the $K-L$ color of AS~353Ba by \citet{Tok04}, 0.56
magnitudes, suggests the possible presence of circumstellar material
around at least one of the AS~353B components.  However, if the $K$-
and $L$-band magnitudes for AS~353Ba are dereddened
by an A$_V$ of 2 magnitudes \citep{Pra03}, the resultant K$-$L color
is 0.44 magnitudes.  For an A$_V$ of 3 magnitudes (this paper),
the dereddened K$-$L color is 0.37 magnitudes.  Therefore, if
circumstellar material is present at all around AS~353Ba, it is likely to
be found at an advanced stage of dispersion.  Possibly the
binarity of AS~353B has played a role in the relatively early
dissolution of its circumstellar disks.  

It has been suggested that AS~353A is
also a close binary \citep{Tok04}.  If true, this would
complicate the idea that binarity drives accelerated disk
evolution, as AS~353A harbors one of the most substantial and active
circumstellar disks known.  \citet{Tok04} find no evidence for a
companion object down to a separation of 0.1$''$, corresponding to
20~AU for a distance of 200~pc to the system.  If we assign
AS~353A a mass of 1.2~M$_{\odot}$ (Figure~\ref{fig:hrd}), the velocity
signature induced by a companion object with a mass
of 0.8~M$_{\odot}$ in a circular orbit is about 7.5~km~s$^{-1}$.  Any higher mass
companion would result in a larger amplitude of radial velocity 
variability in AS~353A, whereas any inclination of the system's orbit out
of the plane of our line of sight, a larger distance to the
system than 200~pc, or a smaller mass companion, would result in
smaller amplitude velocity variations.  Over
an interval of two years, however, no radial velocity variation
greater than 1~$\sigma$ was measured between our three observations.
Our observations, however, may be insensitive to highly eccentric
spectroscopic binary orbits \citep[e.g.][]{Rei00}.

\subsubsection{HBC~294, HBC~681, \& HBC~683}

HBC~294 is best fit by a K7 template spectrum with
$v{\sin}i$=25~km~s$^{-1}$.
This agrees with the spectral type reported in the HBC.
The radial velocity is well-determined, with a standard
deviation of $<$~0.5~km~s$^{-1}$ based on measurements from
three different epochs. HBC~294, also known as V536~Aql, is a
typical cTTs, with Br$\gamma$ emission detected in our $K$-band
observations and H$\alpha$, Pa$\beta$, and optical forbidden
emission lines reported in the literature, indicative of accretion
from a circumstellar disk.  \citet{Whe04} describe evidence for
a bipolar outflow from the system as well.

\citet{Age94} first discovered HBC 294 to be a binary with an 0\farcs52
separation at a position angle of 17\degr.  The unresolved near-IR
excess observed in the system indicates warm dust surrounding at
least one of the stars. The \emph{IRAS} PSC flux measurements of HBC~294
(\emph{IRAS}~19365+1023) reveal even greater mid- to far-IR fluxes
than those of AS~353A at all but 100~\micron~\citep{Wea92}.  It is
unclear whether the circumstellar material is located around one or
both components of this binary.  The lower 100~$\mu$m flux suggests
that circumstellar material is truncated by the binary companion.
The extinction required to deredden HBC~294 to the cTTs locus is
comparable to the average extinction of the sample.

A K5 template with $v{\sin}i$=20~km~s$^{-1}$ provided the best
fit to the spectrum of HBC~681.
The H$\alpha$ emission of HBC~681, EW=35~\AA, is relatively low
for the cTTs in this sample, but still well above the 10~\AA~cTTs
limit.  A clear IR excess is indicated by the location of HBC~681
on the color-color diagram.  We identified a faint potential
companion to HBC~681 at separation $\sim$1\farcs2 and a position
angle of $\sim$195$\degr$.  The $H-K$ color of this object is
$\sim$0.3$-$0.4, typical of an unreddened M6 star, which, at the age of the Aquila
association, would be substellar.  However, the proper motion of this object
must be observed to determine if the two objects are indeed
related as HBC~681 is located in the direction of the galactic
plane, and hence the field is crowded with faint background objects.

For the spectrum of HBC~683, we found a best fit using a K7 template
spectrum with $v{\sin}i$=15~km~s$^{-1}$.
This is later than the spectral type of K1 reported in the HBC.
The 2003 September 08 (UT) order 49 spectrum of HBC~683 is one of the
best for line identification because of the high signal to noise ratio
and low $v{\sin}i$ relative to the rest of the sample.
HBC~683 is a typical cTTs and has the strongest
H$\alpha$ and Br$\gamma$ emission in our sample after AS~353A.
HBC~683 also shows a clear IR excess in the color-color diagram
(Figure~\ref{fig:ccd}).  These properties are all consistent with
its cTTs classification and the presence of an actively accreting
circumstellar disk.

\subsubsection{HBC~682 A \& B}

As a result of the observations presented in this paper, HBC~682 was
discovered to be a visual binary with a separation of $\sim$1\arcsec.
HBC~682A is best fit by a K5 template spectrum with
$v{\sin}i$=50~km~s$^{-1}$, and HBC~682B by a K7 template
spectrum with $v{\sin}i$=30~km~s$^{-1}$. Both of these spectral types
are earlier than that of M0 reported in the HBC.

Although the spectra of both components of HBC~682 
have similar $H$- and $K$-band features, there are striking
rotational and radial velocity differences between the two.
HBC~682A has the broadest absorption lines in our sample, while
HBC~682B is more typical.
A consistent radial velocity was not measured for HBC~682A because of
the large variations between epochs (Table~\ref{table:radvels}).
This may be in part because of the wide and shallow absorption lines
that make precise measurement of radial velocity more difficult.
However, the radial velocity variations measured in 5 different epochs,
were significantly larger than the typical variation for other objects,
e.g., 1$-$2~km~s$^{-1}$.  HBC~682A is therefore a candidate
spectroscopic binary.  We speculate that HBC~682A and B are bound
because of their close proximity.  Verification by component
proper motion measurements and by determination of the center of
mass velocity of the HBC~682A spectroscopic binary is needed,
particularly in that this system is located in projection on to the
galactic plane and hence confusion with background sources is possible.

The angularly unresolved HBC~682 system shows no IR excess,
consistent with its weak H$\alpha$ and lack of Br$\gamma$ emission.
However, the classification of HBC~682 is complicated by the fact that
it has only recently been resolved into a visual binary, and may in
fact be a hierarchical triple.
The components lack separate H$\alpha$ emission measurements and
resolved near-IR photometry, both of which might provide additional
information about their individual circumstellar environments.

\subsubsection{FG~Aql/G3}

An M0 template with $v{\sin}i$=15~km~s$^{-1}$ provided the best
fit to the FG~Aql/G3 system.
The only observations of this object since \citet{Coh79} are from
\emph{2MASS} and \emph{DENIS} (Deep Near Infrared Survey), which
combined provide $I$-, $J$-, $H$-and $K$-band photometry.
No IR excess is detected.
There are as yet no known signatures of youth associated with
FG~Aql/G3; \citet{Coh79} did not detect H$\alpha$ emission from this
star and we did not observe it in the $K$-band.  The $H$-band
spectrum, however, appears highly veiled.  The location within
20\arcsec~of three cluster members and FG~Aql/G3's
measured radial velocity are consistent with cluster membership.
However, further evidence is needed to confirm the youth of this object. 

\subsubsection{HBC~684}

\begin{figure*}
\includegraphics[angle=90,width=6.0in]{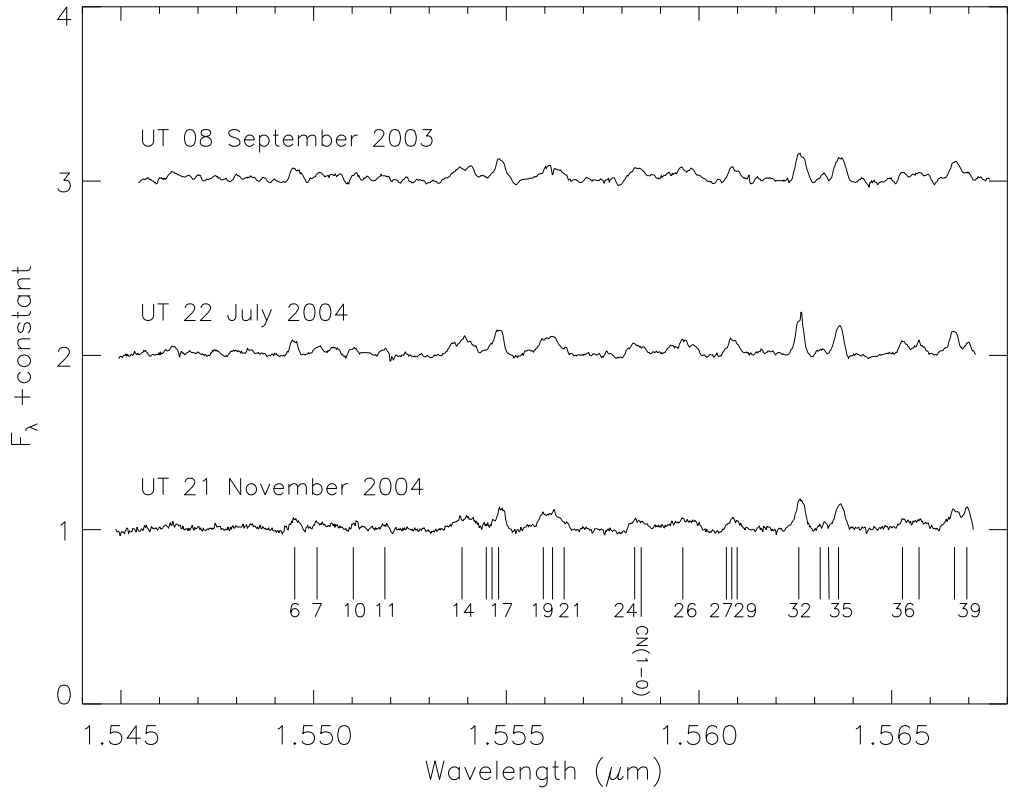}
\caption{Multi-epoch NIRSPEC order 49 spectra of the unusual
  emission-line object HBC~684.  Wavelength and
  species of identified lines are given in Table~\ref{table:lines}.
  All spectra have been
  wavelength-corrected for heliocentric and radial velocity and
  normalized to a continuum level of unity. }
\label{fig:hbc684}
\end{figure*}

\begin{figure*}
\epsscale{1.0}
\includegraphics[angle=0,width=5.0in]{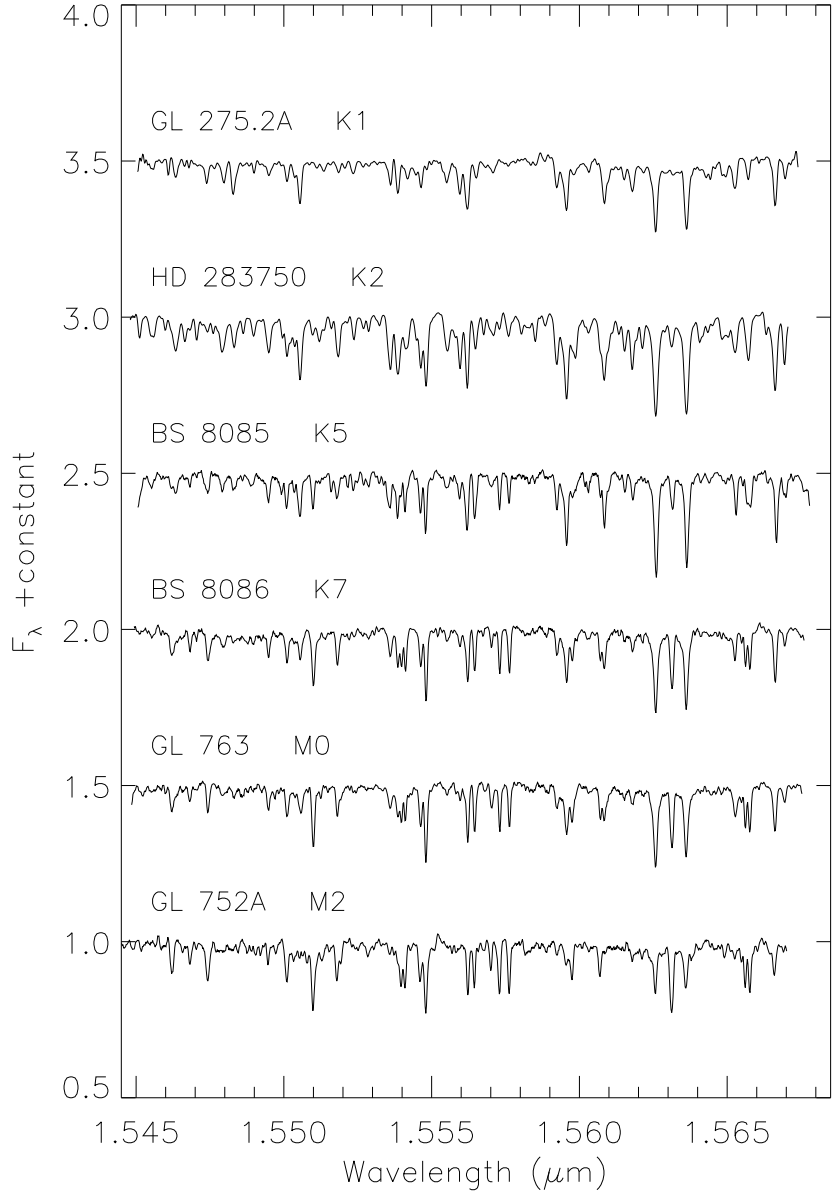}
\caption{Template spectra from \citet{Ben05} used to determine
  spectral type and rotational velocity of each sample object.  These
  spectra have been convolved with a $v{\sin}i$=10~km~s$^{-1}$
  rotation kernel from \citet{Gra92}.
  Spectra have also been corrected for heliocentric and radial
  velocities, normalized to continuum levels of unity, and are
  presented on the laboratory wavelength scale. }
\label{fig:templates}
\end{figure*}

\begin{figure*}
\includegraphics[angle=90,width=6.0in]{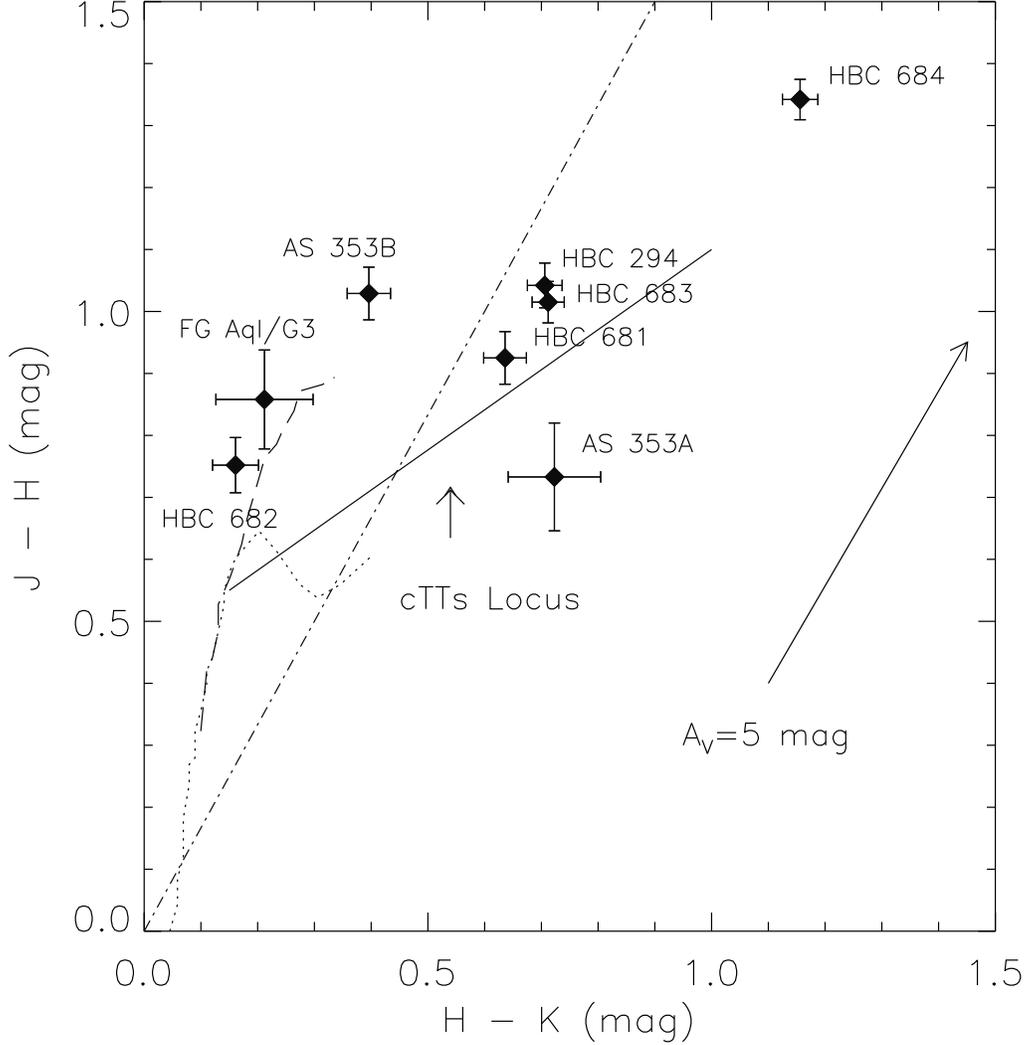}
\caption{$J-H$ versus $H-K$ color-color diagram for the sample objects
  (Table~\ref{table:colors}). Magnitudes were taken from
  \emph{2MASS}.  Error bars represent the propagated photometric errors reported
  in the \emph{2MASS} catalog.  The dash-dotted line separates objects
  with (to the right) and without IR-excess.  The cTTs (solid line), dwarf
  (dotted line), and giant (dashed line) loci are the same as in Figure~3 of
  \citet{Pra03} but transformed in to the \emph{2MASS} magnitude scale
  using equations from \citet[\S4.3]{Car01}.  The effect on observed
  color of 5 magnitudes of visual extinction is represented by the
  arrow (thick line), using the equation derived by \citet[\S3.1]{Pra03}. }
\label{fig:ccd}
\end{figure*}

\begin{figure*}
\includegraphics[angle=90,width=6.5in]{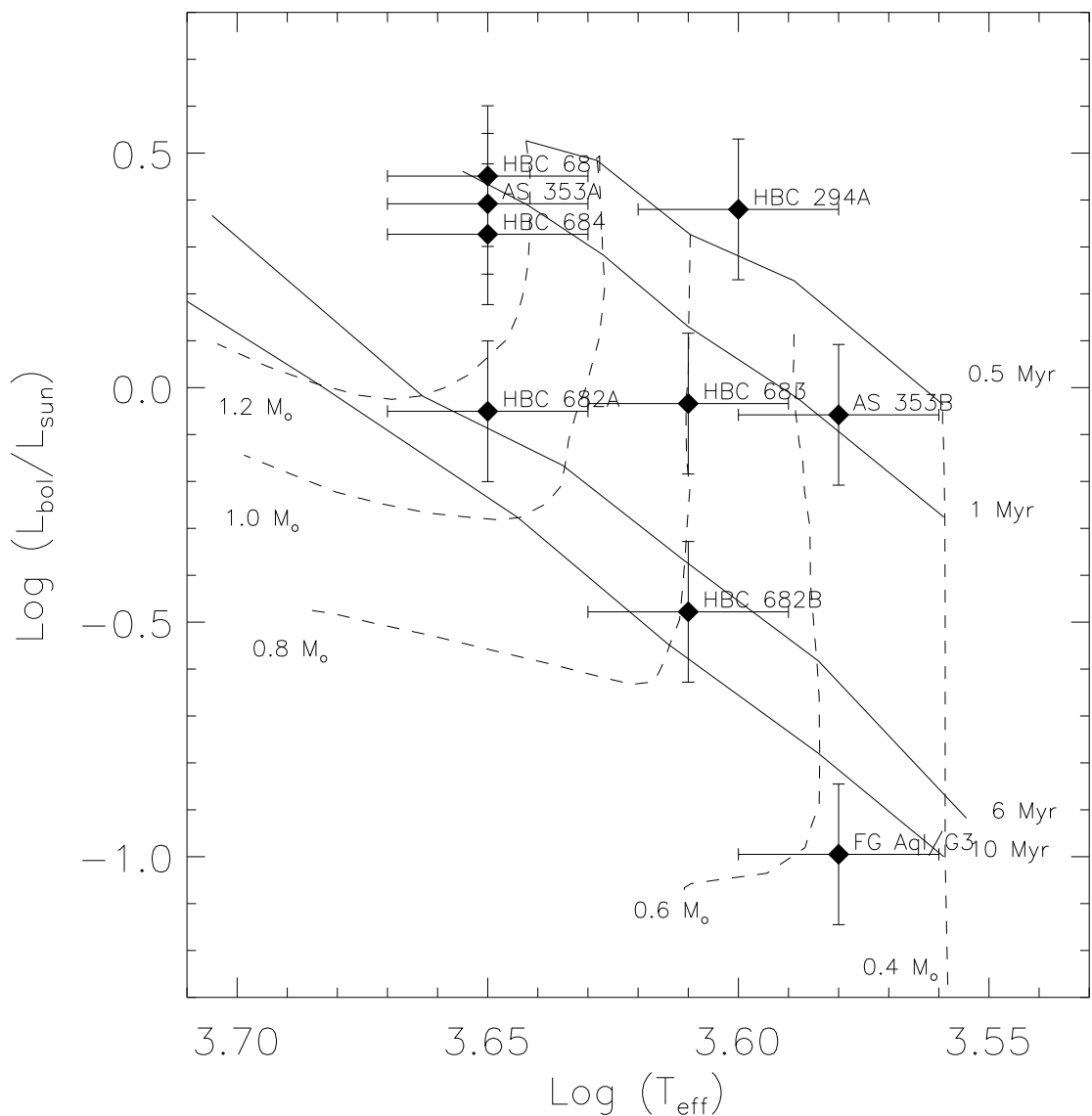}
\caption{H-R diagram of probable Aquila association members using
  theoretical evolutionary tracks from \citet{Pal99}.  
Mass tracks (dotted lines) are 1.2, 1.0, 0.8, 0.6, and 0.4
  solar masses from left to right and isochrones (solid lines) are
  5$\times$10$^{5}$, 10$^{6}$, 6$\times$10$^{6}$, and 10$^{7}$ years from top to bottom.  
Temperatures were taken from \citet{Luh04} based on the spectral types
  listed in Table~\ref{table:results}.  
Although AS 353B and HBC 682 were not resolved in
  \emph{2MASS}, their luminosities have been corrected for
  binarity based on their near-IR flux ratios.}
\label{fig:hrd}
\end{figure*}

The spectra from the three observations of HBC~684 are presented in 
Figure~\ref{fig:hbc684}.  The inverted order 49 spectrum of HBC~684
is best fit by a K5 template spectrum with $v{\sin}i$=40~km~s$^{-1}$.
The source of the emission lines in the spectra of this star is
enigmatic.  An examination of the FITS files revealed that the
emission lines appeared in the raw data.
Furthermore, emission lines were apparent in all orders of the raw
$H$- and $K$-band data.

Although emission lines associated with accretion are common for young
stars, they are usually manifested in strong hydrogen lines
and only occasionally in other species, such as CO and neutral Na
\citep{App89,Eis90,Gre96,Pra03}.
HBC~684 shows only modest H$\alpha$ (11~\AA~in the HBC and
$\sim$13~\AA~in our unpublished visible light spectrum) and Br$\gamma$
emission, while most of the atomic lines in order 49, mainly Fe
but possibly Ti, Si, and Ni as well, all appear in emission.
We have tentatively identified lines of OH and CN also in emission;
see Figure~\ref{fig:hbc684} and Table~\ref{table:lines} for line 
identifications.  In addition to these molecular lines in
order 49, many emission lines probably arising from OH were identified
in orders 46, 47, 48, and 50, and several emission lines from CN in
orders 48 and 50.  All of the typical cTTs is this sample
(HBC~294, HBC~681, and HBC~683) show stronger H$\alpha$ emission,
and all but HBC~294 show stronger Br$\gamma$ emission as
well (Table~\ref{table:EWs}).

\citet{Gre96} reported a rich near-IR emission line spectrum of the
class I young stellar object \emph{IRAS}~04239$+$2436 in Taurus.
Although the resolution was modest (R$\sim$500), over a dozen emission
features were identified in the observed spectral range of 1.1~to~2.5~$\micron$. 
Probable features included several CO bandheads, Br$\gamma$, Br10,
Br11, Br13, Br14, Pa$\beta$, H$_2$, and neutral He.
Only the neutral Ca triplet at $\lambda$=2.26~$\micron$ was observed
in absorption.  We know of no other spectra, however, with low-excitation
atomic and possibly molecular species in emission as they appear in
HBC~684's $H$-band spectra.  

The Palomar Observatory Sky Survey and \emph{2MASS} images show
some nebulosity to the north and east of HBC~684; however, this
diffuse emission was not evident in more recent $H$-band images
(in short, 0.1s exposures) taken with the
NIRSPEC Slit-viewing Camera (SCAM) in 2004 November.
HBC~684 is an \emph{IRAS} source (\emph{IRAS}~19046$+$0508) with PSC
fluxes at 12, 25, 60 and 100~$\micron$ even greater than those of
AS~353A, which indicates a considerable amount of warm dust around
the star.  The near-IR excess of HBC~684 is the strongest in our
sample (Figure~\ref{fig:ccd}).

We suggest four mechanisms that might produce spectra such as
those observed in HBC~684.  First, accretion could be creating an
oblique shock close to the photosphere
of the star, which occurs when the accretion speed equals the escape
velocity of the star.  However, the lack of strong H$\alpha$
emission limits the amount of circumstellar material that could
be accreting from a circumstellar disk.  Furthermore, our H$\alpha$
spectrum shows a clear P~Cygni profile, characteristic of 
outflow, but not infall.

HBC~684's emission lines could also be produced by an inversion layer
resembling the chromosphere of the Sun.  If some blocking
mechanism such as a circumstellar disk, similar to the function of the
Moon during total solar eclipse, could be construed, a phenomenon
resembling a solar flash spectra might be produced
\citep[see][]{Her90}.  Stellar chromospheres are heated by stellar
coronae; the X-ray emission
frequently observed in young stars is thought to originate in the
corona \citep[][e.g.]{Neu95}.  However, the nearest known X-ray
source in the vicinity of HBC~684, 1RXS J190722.4$+$051231
in the \emph{ROSAT} All-Sky Survey Faint Source Catalog,\footnote{The
\emph{ROSAT} All-Sky Survey Faint Source Catalog is queried via the
High Energy Astrophysics Science Archive Research Center(HEASARC)
Browse Main Interface. HEASARC is a service of the Laboratory for
High Energy Astrophysics at NASA/GSFC and the High Energy
Astrophysics Division of the Smithsonian Astrophysical Observatory.}
is 3.2\arcmin~away and is therefore probably not associated with HBC~684. 

A transient impact on the photosphere could provide sufficient energy
to excite the atomic metals that are observed in emission. 
In 1994 July fragments of the comet Shoemaker-Levy~9 plunged
into the atmosphere of Jupiter.
\citet{Cos97} observed the plume of the most energetic impact and
found atomic emission lines from neutral Na, Fe, Ca, Li in the optical spectrum.
Infrared spectra of another impact show molecular band/line emission,
including CH$_4$ and CO(2$-$0) \citep{Kna97}.  Given the few
million year age of this association and the evidence for circumstellar
material around HBC~684, it is plausible that planetessimals are
forming around and may occasionally fall into the star.  Indeed,
the planet plunging phenomenon proposed to account for the high
metallicity observed in extra-solar planetary systems \citep{Arm02} may
be invoked in this case.  The slow motion merger of a giant planet
or brown dwarf could create the requisite energy for the Fe~I
emission lines (on the order of 6$-$7~eV) without exciting significant
hydrogen emission.  Such a merger would require a substantial
circumstellar disk in order to abet angular momentum transfer;
the presence of a disk is supported by the strong near- and mid-IR
excesses observed in this system.  The P~Cygni profile in the
H$\alpha$ emission line implies an outflow of 200~km~s$^{-1}$.
A planet plunging event might stimulate the velocity of such a jet.
The duration of the effects of a merger is not known; we have
observed consistent emission line spectra of HBC~684 on three
separate occasions over a 14 month period.  In 1981, a broad,
strong P~Cygni profile was already present (notes in HBC).

A final possibility we consider is the stimulation of line emission by
fluorescence.  We found that the upper energy level terms for the order 49
Fe~I lines represent too broad a range of states for the production
of all the $H$-band lines to be consistent with fluorescence
originating in one particular pumping line.
Furthermore, all the lines expected in a normal K5 star are
present in the infrared spectra of HBC~684, except in emission.
Compare this, for example, to HBC~682B, which has a similar spectral
type and value of $v{\sin}i$.  HBC~684's emission lines also display similar line
ratios as the corresponding absorption lines in a template
spectrum or in HBC~682B.  Some of the unusual features in the
HBC~684 spectra may be the result of fluorescence, but we are
skeptical that this is the explanation for the entire set of
emission lines observed.

\section{\sc Conclusion}

This paper presents the results of multi-epoch observations of a
sample of nine young, late-type stars in the Aquila star-forming region. 
Sample objects were selected from \citet{Her88} based on their
proximity to AS~353A and classification as a wTTs or cTTs.
The high-resolution (R$\sim$30,000) $H$-and $K$-band observations were
made between 2002 July and 2004 November with NIRSPEC on Keck~II.
The initial goal of this project was to examine one $H$-band order of
the multiple-epoch spectra for radial velocity variations in order to
detect spectroscopic binaries.
One radial velocity variable object was discovered and absolute
radial velocities were measured for the remaining eight objects.
Circumstellar properties of the stars were also examined via
high-resolution $K$-band spectra and \emph{2MASS} photometry.

Eight of the nine sample objects appear to be members of the same
association because of similar radial velocities.
The spatial extent of the known members, using our adopted
distance of 200$\pm$30~pc, is 32$\pm$5~pc by 58$\pm$9~pc.
The ages of the observed objects exhibit a large scatter, with the
youngest, most active members all of age $\sim$1$-$2~Myr.  Two systems
appear to have ages between 6 and 12 Myr, and the oldest of these
(FG~Aql/G3) has not been confirmed to show signatures of youth.
Therefore, the Aquila association of stars is apparently one of the
youngest nearby star-forming regions.
The visual extinction to the association is estimated to be about 1$-$2 magnitudes.
 
The circumstellar environments of associations members are varied,
ranging from extremely active (AS~353A, whose winds drive HH 32) to
nearly devoid of circumstellar material (AS~353B, HBC~682, and
FG~Aql/G3, which all lack IR excess and strong H$\alpha$ emission),
with several typical cTTs in between (HBC~294, HBC~681, and HBC~683).
The source of the atomic and molecular emission lines in the spectra
of HBC~684 is as yet unknown, but may be caused by an inversion
layer, increased accretion, fluorescence, or a transient impact.
For the seven systems in this study, HBC~681, HBC~682, HBC~683,
HBC~684, HBC~294, FG~Aql/G3, and AS~353, there is a high total
number of stars, thirteen at least, yielding $\sim$2 stars per
system.  This multiplicity is significantly higher than even that
of the Taurus region ($\sim$1.6 stars per system; \citealt{Sim95}),
although it applies to only a small sample and may be
significantly impacted by observational biases.
The extremely interesting stars identified to date in this association
certainly warrant further investigation.

Much work remains to characterize this unusual star-forming region.
The most pressing task is to confirm the existence of a larger association
by searching for other young stars in the region and identifying
radial velocities and proper motions of the new members.
Obtaining a more accurate distance determination to the association
will also be critical to unraveling the age-distance discrepancy.
For the non-thermal radio source AS~353B, this could be accomplished
with VLBI \citep[e.g.][]{Loi05}.
Once more complete membership is determined, the Aquila
star-forming region can be characterized in terms of mass function,
multiplicity fraction, and other parameters. 
Further observations in the millimeter, sub-millimeter,
infrared, radio, and X-ray will also
provide valuable information about the circumstellar environments of
these objects, which will improve our understanding of key
phases of pre-main-sequence stellar evolution. 

\section{\sc Acknowledgments}

We thank the expert observing assistants and
staff of the Keck Observatory for their capable help.
Insightful discussions with C. McCabe, D. Hunter, M. Simon,
S. Kenyon, E. Becklin, M. Jura, B. Zuckerman, S. Strom, G. Herbig, and B. Reipurth improved the
presentation of this paper.  We are grateful to S. Zoonematkermani for
providing some of the IDL procedures used in the analysis, to
C. McCabe and G. Duch\^ene for obtaining several spectra
for us in 2004 May, to M. Huerta for speedy reduction of the
AS~353A and HBC~684 H$\alpha$ spectra taken at McDonald
Observatory in 2005 November, and to Q. Konopacky for several NIRC2 images.
We thank the anonymous referee for helpful comments.
L. Prato was supported in part by NSF grant AST 04-44017.
Data presented herein were obtained at the W. M. Keck Observatory,
which is operated as a scientific partnership among the California
Institute of Technology, the Universities of California, and the
National Aeronautics and Space Administration. The observatory was
made possible by the generous financial support of the W. M. Keck
Foundation. The authors wish to recognize and acknowledge the very
significant cultural role and reverence that the summit of Mauna Kea
has always had within the indigenous Hawaiian community. We are most
fortunate to have the opportunity to conduct observations from this mountain.
This work made use of the SIMBAD reference database, the
VizieR Service, the NIST Atomic Spectra Database, the NASA
Astrophysics Data System, and the NASA/IPAC Infrared Service Archive.

\clearpage
\begin{deluxetable}{lllllrr}
\tablecaption{\sc The Sample}
\tabletypesize{\small} \tablewidth{0pt}
\rotate
\tablehead{
\colhead{Primary } & 
\colhead{Other } & 
\colhead{R.A.\tablenotemark{a} } & 
\colhead{Dec.\tablenotemark{a} } &
\colhead{$l,~b$\tablenotemark{a}} &
\colhead{$H$-band} &
\colhead{EW(H$\alpha$)\tablenotemark{b}  } \\ 
 \colhead{Name } & 
\colhead{Name } & 
\colhead{(J2000.0) } & 
\colhead{(J2000.0) } &
\colhead{(\degr,~\degr) } &
\colhead{magnitude\tablenotemark{a}} &
\colhead{(\AA) }
}
\startdata

HBC~681  & FG~Aql/G1 & 19 02 22.20 & $-$05 36 20.4 & 29.17, $-$4.98 & 9.429 & 35 \\
FG~Aql/G3 & \emph{2MASS} 190222.6$-$053522  & 19 02 22.60 & $-$05 36 22.1 & 29.17, $-$4.98 & 12.019 & \nodata \\
HBC~682\tablenotemark{c}  & FG~Aql/G2 & 19 02 22.84 & $-$05 36 15.8 & 29.18, $-$4.98 & 9.492 & 2.3 \\
HBC~683  & FH Aql & 19 02 23.22 & $-$05 36 37.3 & 29.17, $-$4.98 & 10.380 & 105 \\
HBC~684\tablenotemark{d}  & \emph{IRAS}~19046$+$0508 & 19 07 09.87 & $+$05 13 10.8 & 39.37, $-$1.11 & 8.555 & 11 \\
AS~353A  & HBC~292 &19 20 30.99  & $+$11 01 54.7 & 46.05, $-$1.33 & 9.161 & 150 \\
AS~353B\tablenotemark{e}  & HBC~685 & 19 20 31.04 & $+$11 01 49.1 & 46.05, $-$1.33 & 9.154 & 4.4 \\
HBC~294\tablenotemark{f}  & V536 Aql & 19 38 57.41 & $+$10 30 16.1 & 47.75, $-$5.57 & 8.102 & 52 \\
\enddata
\tablenotetext{a}{R.A. and Dec. from \emph{2MASS}; Galactic coordinates from SIMBAD}
\tablenotetext{b}{From \citet{Her88}}
\tablenotetext{c}{1\arcsec~visual binary}
\tablenotetext{d}{Some confusion exists in the literature because of
  the identification of HBC~684 as WL 22 in several papers,
  including \citet{Her88}, \citet{Leva88}, and \citet{Wea92}. In the
  SIMBAD database, WL 22 refers to a T~Tauri star in Ophiucus.}
\tablenotetext{e}{0\farcs24 binary}
\tablenotetext{f}{0\farcs52 binary}
\label{table:coords}
\end{deluxetable}

\clearpage

\begin{deluxetable}{llcccc}
\tablecaption{\sc Observing Log}
\tabletypesize{\footnotesize} \tablewidth{0pt} 
\tablehead{
\colhead{Date } & 
\colhead{Object } & 
\colhead{Filter } & 
\colhead{Integration Time } &
\colhead{Airmass } &
\colhead{SNR } \\

\colhead{(UT) } & 
\colhead{ } & 
\colhead{ } & 
\colhead{(s) } &
\colhead{ } &
\colhead{ }

}
\startdata 
2002 July 18 & AS~353A & N5 & 300 & 1.03  & 170 \\ 
  & AS~353B & N5 & 300 & 1.04 & 200 \\
2003 August 10 & HBC~681 & N5 & 300 & 1.11 & 350  \\
  & FG~Aql/G3 & N5 & 220 & 1.11 & 250 \\
  & HBC~682 (A \& B) & N5 & 180 & 1.11 & 280\\ 
2003 September 07 & HBC~681 & K & 240 & 1.12  & 150 \\
  & HBC~682 (A \& B) & K & 200 & 1.13 & 180 \\
  & HBC~683  & K & 240 & 1.16 & 170 \\
  & HBC~684  & K & 120 & 1.11 & 170 \\
  & HBC~294  & K & 120 & 1.05 & 180 \\
2003 September 08 & HBC~681 & N5 & 240 & 1.13 & 250 \\
  & HBC~682 (A \& B) & N5 & 240 & 1.19 & 270 \\
  & HBC~683 & N5 & 220 & 1.40 & 250 \\
  & HBC~684 & N5 & 300 & 1.49 & 280 \\
  & HBC~294 & N5 & 300 & 1.40 & 420 \\
2004 May 27 (AO) & HBC~681 & N5 & 300 & 1.25 & 210 \\
  & AS~353A & N5 & 300 & 1.02 & 170\\
2004 May 28 (AO) & HBC~682A & N5 & 300 & 1.22 & 250 \\
2004 July 22  & HBC~682A & N5 & 240 & 1.28 & 300 \\
  & HBC~681 & N5 & 180 & 1.36 & 300 \\
  & HBC~684 & N5 & 180 & 1.30 & 340 \\
  & AS~353A & N5  & 300 & 1.05 & 210 \\
  & HBC~294 & N5 & 120 & 1.06 & 320 \\
2004 November 21 & HBC~683 & N5 & 240 & 1.56 & 80 \\
  & HBC~684 & N5 & 180 & 1.53 & 330 \\
  & HBC~294 & N5 & 120 & 1.39 & 270 \\
2004 November 22 & HBC~682 (A \& B) & N5 & 180 & 1.57 & 210 \\
  & AS~353A  & K & 240 & 1.49  & 340  \\
\enddata
\label{table:obslog}
\end{deluxetable}

\begin{deluxetable}{llrccc}
\tablecaption{\sc NIRSPEC Instrument Settings}
\tabletypesize{\small} \tablewidth{0pt} 
\tablehead{
\colhead{Filter } & 
\colhead{Date } & 
\colhead{Echelle } & 
\colhead{Cross-dispersion } &
\colhead{Slit Size } &
\colhead{Spectral } \\
\colhead{ } & 
\colhead{(UT) } & 
\colhead{Angle (\degr) } & 
\colhead{Angle (\degr) } &
\colhead{(Arcseconds) } &
\colhead{Resolution }
}

\startdata 
N5  & 2002 Jul. - 2004 Nov. & 63.04 & 36.30 & 0.288 $\times$ 24 & 30,800\tablenotemark{a} \\
N5 (AO) & 2004 May 27 \& 28  & 63.04 & 36.30 & 0.027 $\times$ 2.26 & 35,900\tablenotemark{b} \\
K & 2003 September 07 & 62.15 & 35.59  & 0.288 $\times$ 24  & 24,500\tablenotemark{c} \\
K & 2004 November 22 & 62.84 & 35.59 & 0.288 $\times$ 24  & 29,000\tablenotemark{c} \\
\enddata
\tablenotetext{a}{Order 49; calculated using the FWHM of an OH night sky line and averaged over all frames.}
\tablenotetext{b}{Order 49; calculated using the FWHM of an arc lamp line and averaged over both nights.}
\tablenotetext{c}{Order 35; calculated using the FWHM of an OH night sky line and averaged over all frames.} 
\label{table:inst}
\end{deluxetable}

\begin{deluxetable}{lrrc}
\tablecaption{\sc Measured Brackett Strengths}
\tabletypesize{\small} \tablewidth{0pt}
\tablehead{
\colhead{Object } & 
\colhead{Obs. Date } &
\colhead{EW\tablenotemark{a}} & 
\colhead{Line Flux\tablenotemark{b}} \\
\colhead{ } & 
\colhead{(UT) } &
\colhead{(\AA) } & 
\colhead{(W$-$m$^{-2}\times10^{-16}$)}
}
\startdata
         & Br$\gamma$ & & \\
\hline
HBC~681  & 2003 September 07 & $-$5 & 0.6$\pm$0.1 \\
HBC~682A & 2003 September 07 & 1 & \nodata\tablenotemark{d} \\
HBC~682B & 2003 September 07 & 1 & \nodata\tablenotemark{d} \\
HBC~683  & 2003 September 07 & $-$5 & 0.3$\pm$0.1 \\
HBC~684  & 2003 September 07 & $-$3 & 1.4$\pm$0.5 \\
HBC~294  & 2003 September 07 & 0\tablenotemark{c} & 0.2$\pm$0.4 \\
AS~353A  & 2004 November 22 & $-$22  & 3.8$\pm$0.2 \\
\hline
         & Br16 & & \\
\hline
AS~353A & 2002 July 18 & $-$5 & 1.6$\pm$0.3 \\
AS~353A & 2004 May 27 & $-$7 & 2.2$\pm$0.3 \\
AS~353A & 2004 July 22 & $-$4 & 1.4$\pm$0.3 \\
\enddata

\tablenotetext{a}{Br$\gamma$ or Br 16 equivalent widths.  Negative values indicate emission and positive values absorption.  Uncertainties are $\sim$~1~\AA.}
\tablenotetext{b}{Flux integrated over emission feature and corrected
  for continuum emission.}
\tablenotetext{c}{Br$\gamma$ emission was observed in
  the spectrum of HBC~294, but the measured value was less than 0.5~\AA.}
\tablenotetext{d}{~Line flux was not calculated for absorption lines.}
\label{table:EWs}
\end{deluxetable}

\begin{deluxetable}{lc}
\tablecaption{\sc Spectral Type and $v{\sin}i$ Templates\tablenotemark{a}}
\tabletypesize{\small} \tablewidth{0pt}
\tablehead{
\colhead{Object } & 
\colhead{Spectral Type } 
}
\startdata
GL 275.2A & K1 \\
HD 283750 & K2 \\
BS 8085 & K5 \\
BS 8086 & K7 \\
GL 763 & M0 \\
GL 752A & M2 \\
\enddata
\tablenotetext{a}{Template spectra obtained from C. Bender, private communication.}
\label{table:templates}
\end{deluxetable}

\begin{deluxetable}{lccccc}
\tablecaption{\sc Spectral Type, $v{\sin}i$, and Radial Velocity}
\tabletypesize{\small} \tablewidth{0pt}
\tablehead{
\colhead{Object } & 
\colhead{Spectral Type } & 
\colhead{$v{\sin}i$ } &
\colhead{Radial Velocity } & 
\colhead{Standard Deviation }  &
\colhead{No. } \\
\colhead{ } & 
\colhead{ } & 
\colhead{(km~s$^{-1}$) } &
\colhead{(km~s$^{-1}$) } & 
\colhead{(km~s$^{-1}$) } &
\colhead{Spectra\tablenotemark{a} }  \\
\colhead{(1) } & 
\colhead{(2) } & 
\colhead{(3) } &
\colhead{(4) } & 
\colhead{(5) }  &
\colhead{(6) } 
}
\startdata
HBC~681  & K5 & 20 & $-$7.4 & 0.9 & 4 \\
FG~Aql/G3 & M0 &  15 & $-$8.9  & \nodata\tablenotemark{b} & 1 \\
HBC~682A  & K5 & 50  & \nodata\tablenotemark{c} & 5.6 & 5 \\
HBC~682B  & K7 & 30  & $-$6.8 & 1.5 & 3 \\
HBC~683  & K7 & 15 & $-$6.8  & 0.8 & 2 \\
HBC~684  & K5 & 40 & $-$7.2 & 2.2 & 3 \\
AS~353A  & K5 & 10 & $-$11.4 &  1.1 & 3 \\
AS~353B  & M0 & 20 & $-$10.7 & \nodata\tablenotemark{b} & 1 \\
HBC~294  & K7 & 25 & $-$9.8 & 0.3  & 3 \\

\enddata
\tablenotetext{a}{Number of spectra used in radial velocity analysis.}
\tablenotetext{b}{Only one observation.}
\tablenotetext{c}{Significant variations in radial velocity between epochs $-$ see Table~\ref{table:radvels}.}
\label{table:results}
\end{deluxetable}

\begin{deluxetable}{llr}
\tablecaption{\sc Radial Velocity Measurements}
\tabletypesize{\small} \tablewidth{0pt}
\tablehead{
\colhead{Object } &
\colhead{Obs. Date } &
\colhead{Radial Velocity } \\
\colhead{ } &
\colhead{(UT) } &
\colhead{(km~s$^{-1}$) }
}
\startdata
HBC~681  & 2003 August 10 & $-$8.1 \\
         & 2003 September 08 & $-$7.7   \\
         & 2004 May 27 & $-$6.4   \\
         & 2004 July 22 & $-$7.6   \\
\hline
FG~Aql/G3 & 2003 August 10 & $-$9.4 \\
\hline
HBC~682A & 2003 August 10 & -3.0    \\
         & 2003 September 08 & $-$13.2   \\
         & 2004 May 28 & $-$6.9   \\
         & 2004 July 22 & $-$6.8    \\
         & 2004 November 22 & $-$2.1   \\
\hline
HBC~682B  & 2003 August 10 & $-$5.9 \\
         & 2003 September 08 & $-$5.5   \\
         & 2003 November 22 & $-$8.5   \\
\hline
HBC~683  & 2003 September 08 & $-$7.4 \\
         & 2004 November 21 & $-$6.2  \\
\hline
HBC~684  & 2003 September 08 & $-$5.5 \\
         & 2004 July 22 & $-$7.2  \\
         & 2004 November 21  & $-$9.8   \\
\hline
AS~353A  & 2002 July 18 & $-$12.1   \\
         & 2004 May 27 & $-$10.1   \\
         & 2004 July 22 & $-$12.1   \\
\hline
AS~353B  & 2002 July 18 & $-$10.7 \\
\hline
HBC~294  & 2003 September 08 & $-$9.4   \\
         & 2004 July 22 & $-$9.8  \\
         & 2004 November 21 & $-$9.8   \\
\enddata
\label{table:radvels}
\end{deluxetable}

\begin{deluxetable}{lccc}
\tablecaption{\sc Colors and Extinction}
\tabletypesize{\small} \tablewidth{0pt}
\tablehead{
\colhead{Object } & 
\colhead{$J-H$\tablenotemark{a} } & 
\colhead{$H-K$\tablenotemark{a} } & 
\colhead{Visual Extinction\tablenotemark{b} } \\
\colhead{ } & 
\colhead{ } & 
\colhead{ } & 
\colhead{(mag)}  
}

\startdata
HBC~681  & 0.925 & 0.636 &  ~0.1 \\
FG~Aql/G3 & 0.858 & 0.212 & ~2.7 \\
HBC~682  & 0.752 & 0.161 &  ~1.6 \\
HBC~683  & 1.015 & 0.712 &  ~0.7 \\
HBC~684  & 1.342 & 1.156 &  ~1.6 \\
AS~353A  & 0.733 & 0.723 & $-$3.3 \\
AS~353B  & 1.029 & 0.396 &  ~3.5 \\
HBC~294  & 1.042 & 0.706 &  ~1.1 \\
\enddata
\tablenotetext{a}{Colors calculated from \emph{2MASS} photometry.}
\tablenotetext{b}{Extinction calculated using Equation 2, \S5.3.}
\label{table:colors}
\end{deluxetable}

\begin{deluxetable}{rlrr}
\tablecaption{\sc Multiplicity}
\tabletypesize{\small} \tablewidth{0pt}
\tablehead{
\colhead{Object } & 
\colhead{Multiplicity } &
\colhead{Separation } & 
\colhead{Position Angle } \\
 \colhead{ } & 
\colhead{ } &
\colhead{(\arcsec) } &  
\colhead{(\degr) }
}

\startdata
HBC~681   & Possible low-mass companion & $\sim$1.2 & $\sim$195  \\
FG~Aql/G3 & Single & \nodata & \nodata \\
HBC~682   & Binary, probable triple & $\sim$1 & $\sim$160 \\
      A   & possible SB & \nodata & \nodata \\
HBC~683   & Single & \nodata & \nodata \\
HBC~684   & Single & \nodata & \nodata \\
AS~353    & Triple &  5.6 & 174 \\
      B   & Binary & 0.24 & 107 \\
HBC~294   & Binary & 0.52 & 17 \\
\enddata
\label{table:mult}
\end{deluxetable}

\begin{deluxetable}{llc}
\tablecaption{\sc Identified Absorption Lines}
\tabletypesize{\small} \tablewidth{0pt}
\tablehead{
\colhead{Wavelength } & 
\colhead{Species } &
\colhead{Label } \\
\colhead{($\micron$) } & 
\colhead{ } &
\colhead{ }
}
\startdata
1.54562 & Fe  & 0 \\
1.54635 & Fe  & 1 \\
1.54740 & S  & 2 \\
1.54797 & Fe/S\tablenotemark{a}  & 3\\
1.54829 & S/Fe\tablenotemark{a} & 4 \\
1.54897 & Fe & 5 \\
1.54951 & Fe & 6 \\
1.55009 & Fe & 7 \\
1.55036 & Fe & 8 \\
1.55053 & Fe\tablenotemark{a} & 9 \\
1.55103 & C & 10 \\
1.55185 & Fe & 11 \\
1.55236 & Fe & 12 \\
1.55360 & Fe & 13 \\
1.55385 & Fe & 14 \\
1.55448 & Ni & 15 \\
1.55463 & Fe & 16 \\
1.55480 & Ti & 17 \\
1.55552 & Fe\tablenotemark{a} & 18 \\
1.55596 & Ni & 19 \\
1.55620 & Si & 20 \\
1.55650 & Fe & 21 \\
1.55710 & Fe & 22 \\
1.55760 & Fe & 23 \\
1.55833 & Fe & 24 \\
1.55925 & Fe & 25 \\ 
1.55958 & Fe & 26 \\
1.56071 & Ti & 27 \\ 
1.56085 & Fe & 28 \\
1.56099 & Ni & 29 \\
1.56154 & Fe & 30 \\ 
1.56179 & Fe & 31 \\
1.56259 & Fe & 32 \\
1.56314 & OH & 33 \\
1.56336 & Fe\tablenotemark{a} & 34 \\
1.56362 & Fe & 35 \\
1.56528 & Fe & 36 \\
1.56571 & Fe & 37 \\
1.56663 & Fe & 38 \\
1.56695 & Fe & 39 \\

\enddata
\tablecomments{~Lines were identified using \citet{Liv91} and \citet{Wal01}.}
\tablenotetext{a}{Blended}
\label{table:lines}
\end{deluxetable}

\clearpage

\end{document}